# Semantic Interior Mapology: A Toolbox For Indoor Scene Description From Architectural Floor Plans

Viet Trinh
Computer Science and Engineering
University of California, Santa Cruz
vqtrinh@ucsc.edu

Roberto Manduchi
Computer Science and Engineering
University of California, Santa Cruz
manduchi@soe.ucsc.edu

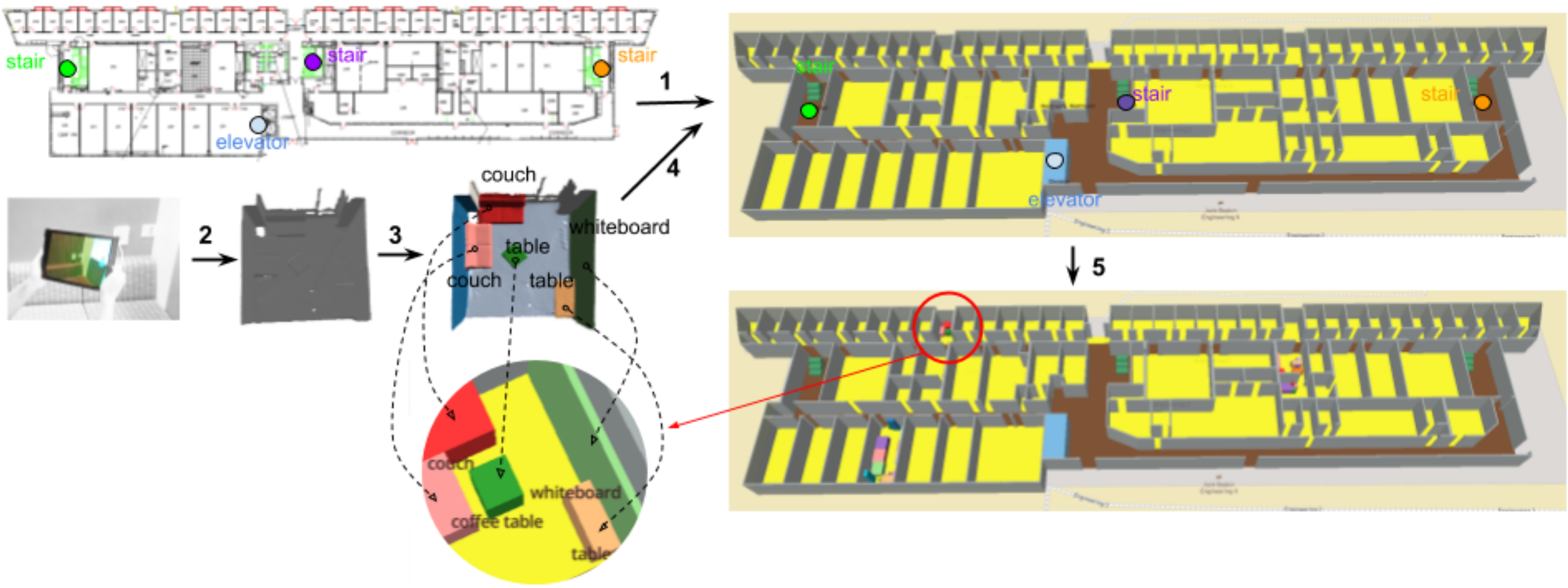

**Figure 1: The SIM toolbox workflow. (1) A floor plan is manually traced and visualized as an extruded 3D object on OpenStreetMap, with some of its fixtures labeled. (2) A room is scanned with an RGB-D sensor. (3) Objects of interest, such as furnitures, are segmented using a semi-automatic process. (4) The 3D room scan is registered with the floor plan, and (5) objects are automatically populated on to the map at the correct locations.**

## ABSTRACT

We introduce the Semantic Interior Mapology (SIM) toolbox for the conversion of a floor plan and its room contents (such as furnitures) to a vectorized form. The toolbox is composed of the Map Conversion toolkit and the Map Population toolkit. The Map Conversion toolkit allows one to quickly trace the layout of a floor plan, and to generate a GeoJSON file that can be rendered in 3D using web applications such as Mapbox. The Map Population toolkit takes the 3D scan of a room in the building (acquired from an RGB-D camera), and, through a semi-automatic process, populates individual objects of interest with a correct dimension and position in the GeoJSON representation of the building. SIM is easy to use and produces accurate results even in the case of complex building layouts. The SIM toolbox is available at https://sim.soe.ucsc.edu.

## KEYWORDS

3D indoor map, 3D content creation, 3D web application for floor plan analysis

## 1 INTRODUCTION

Interactive 3D visualization of building interiors provides enhanced experience of spatial exploration with respect to traditional static maps. Using mapping platforms such as Mapbox[1], 3D pop-up environments can be easily rendered on top of generic 2D maps from web applications such as Google Map or OpenStreetMap (Fig. 1). This type of 3D rendering may afford more intuitive and engaging access to complex building layouts, and may enable interactive features such as displaying a selected floor of a building, or activating groups of features on different levels of detail.

In order to display building interiors by means of 3D interactive tools, it is first necessary to convert available spatial data into an appropriate vectorized format. While most modern building have detailed CAD floor plans (e.g. in dwg or dwf format), this data is normally not accessible to the mapper. Floor plans, when they are available, are only accessible in an image (e.g., JPEG) or PDF format. Computer vision algorithms for the automatic conversion, from the raster to the vectorized form of floor plan images, have been demonstrated, but these algorithms are not universally applicable because of the wide variety of graphical representations used to draw the floor plans. And while companies such Google and Apple are actively acquiring digital representation of interiors of public spaces, and some of these are already available for visualization in their map applications, this data is proprietary and not available to the public. It is also important to observe that floor plans typically contain information only at the level of walls and openings, such as doors. Rarely do they represent smaller-scale features such as fixtures or furnitures. Yet, when available, these features could make for a richer visualization, and could convey useful spatial information. An example is given in the the lower left inset of Fig. 1, wherein the room has been populated with items such as tables, couches, and a whiteboard, which have been correctly geo-registered with the building.

[1]Mapbox. https://www.mapbox.com/

This paper introduces Semantic Interior Mapology (SIM), a toolbox with two main components: (1) The *Map Conversion* toolkit, which is designed to easily convert floor plans into a digital format amenable to interactive visualization; (2) The *Map Population* toolkit, which allows one to add small-scale items that are not present in the original floor plan (Fig. 2). More specifically:

- The Map Conversion toolkit, described in Sec. 3.1, is a web application with an intuitive interface. It allows one to quickly and accurately trace a floor plan from an image of it, and generate a vectorized map. Complex building layouts, like the one shown in Fig. 2, can be traced in just a few minutes. A "semantic" representation of the building layout is saved in a intermediate file format called *sim*, which can then be easily converted into other formats (e.g. GeoJSON, KML, IndoorGML).
- The Map Population toolkit, described in Sec. 3.2, starts from a 3D scan of an environment (e.g., a room). A semi-automatic procedure enables segmentation of the visible surfaces into objects of interest, such as furnitures. This spatial data is then geo-registered with the GeoJSON representation of the same environment generated by the Map Conversion toolkit, and used to "populate" the interior by adding the desired items in the same GeoJSON file. The 3D scans can be obtained using off-the-shelf RGB-D cameras and stitching softwares, such as Occipital's Structure Sensor[2].

To render a floor plan's 3D map view, we employ the MapboxGL JS[3] engine, a location data platform. Geodetic features stored in a GeoJSON file are shown as extruded 3D objects on OpenStreetMap, which can be accessed and interacted with from a regular web browser. This paper is organized as follows. After reviewing the related work in Sec. 2, we proceed with the description of the Map Conversion toolkit in Sec. 3.1 and the Map Population toolkit in Sec.3.2. The conclusions of our work are in Sec. 4.

[2]Structure Sensor. https://structure.io/
[3]MapboxGL JS. https://docs.mapbox.com/mapbox-gl-js/

## 2 RELATED WORK

Early works in the analysis of floor plans focused on the interactive conversion of a 2D raster image into a 3D model [6] [14]. The *ScanPlan* project [17] used the Hough transform for the detection of walls and doors, under the assumption of convex room shape. The algorithm of Ahmed *et al.* [1] detected and labeled rooms based on geometric reasoning involving analysis of the line thickness in a high-resolution image of a small floor plan, typically containing 4 to 5 rooms. De las Heras *et al.* [4] proposed to detect walls based on specific assumptions (i.e. walls are drawn as parallel lines in repetitive patterns that are well distributed in the floor plan). Similarly, the algorithm of [11] recognized walls by determining parallel lines separated by a defined distance. Jang *et al.* [12] used a neural network (U-net) to pre-process the floor plan image and extract a skeleton of walls.

Other recent work employs graph-based algorithms to detect a room's boundary [20] [18], trains a neural network for the task of pixel-wise wall segmentation [5] [15] [19], or reasons about a floor plan's structure from a mobile device's inertial data through crowd-sourcing [22] [13]. Specifically, [10] [9] leveraged crowd-sensed data from mobile users to obtain the spatial relationship between adjacent objects to complete a floor plan reconstruction. Liu *et al.* [16] proposed a neural network, called *FloorNet*, that reasons local spatial information based on the point density captured from smartphones.

Unfortunately, automatic methods for the extraction of room layouts often fail in the case of complex floor plan images. For example, the state-of-the-art algorithm of Liu *et al.* [15] only reaches an accuracy of 85% for room segmentation. In addition, these algorithms usually break down in the case of large and complex layouts such as those considered in this paper (e.g., office buildings), and are unable to correctly detect diagonal walls or nested rooms. In contrast, our proposed Map Conversion toolkit allows for an accurate and fast manual tracing of even very complex building layouts.

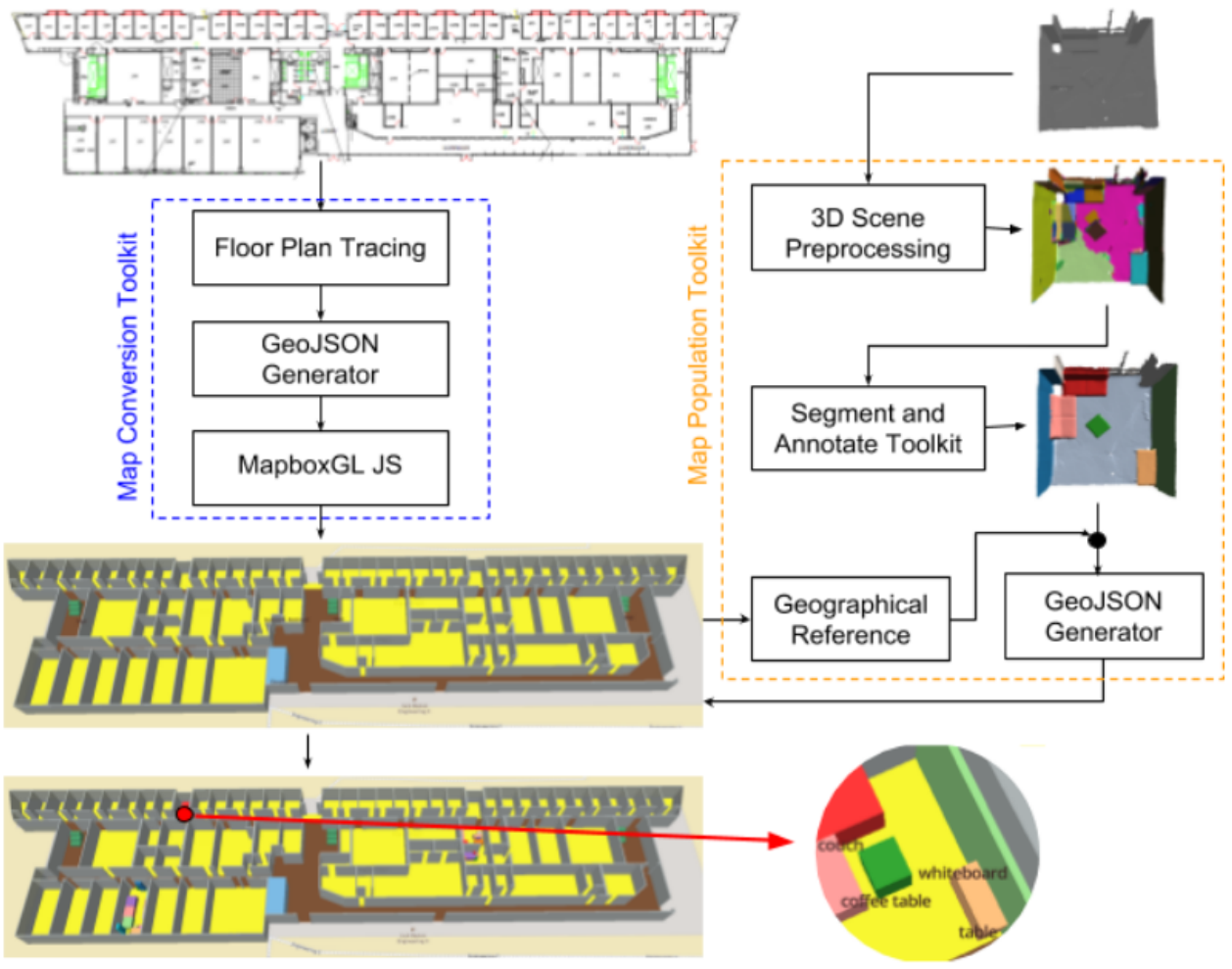

**Figure 2: The SIM toolbox pipeline includes the Map Conversion and the Map Population toolkits.**

# 3 THE SIM TOOLBOX

## 3.1 Map Conversion Toolkit

*3.1.1 Floor Plan Tracing.* The Map Conversion toolkit is a web interface that enables a quick and accurate tracing of a floor plan from an image of it. Our tracing interface exploits the fact that most floor plans have straight walls that intersect at 90 degrees, meaning that most walls can only have one of two orientations. Note that the toolkit also supports less common situations with walls at arbitrary orientation. The floor plan should be oriented such that the main wall orientations are parallel to the screen axes (Fig. 3).

Rather than tracing wall segments by selecting endpoints (as with other web applications such as Google My Map or Mapbox Studio), the Map Conversion toolkit lets the user define a grid of horizontal and vertical lines, where each line overlaps with a segment in the floor plan representing a wall. The user simply Shift-clicks on a segment to generate a line with the desired orientation. Note that, in typical layouts, the same line may contain multiple disjoint wall segments that happen to be co-planar. This strategy is very convenient in the case of repetitive layouts, as it reduces the number of required input selections, and ensures that co-planar walls are traced by segments that are correctly aligned with each other (Fig. 4 (a)(b)). In the case of diagonal (but still planar) wall segments, the user should add two properly oriented "ghost wall" lines, crossing an actual wall segment at the desired corner (i.e., at an endpoint of the diagonal wall segment). Non-planar walls are not currently supported by the toolkit. Lines can be added with simple Shift-clicks, and removed (in the case of a mistake) with Alt/Cmd-clicks. Once all visible wall segments have been covered by lines, the user may click on the *Compute Corners* button. This triggers computation and display, in the form of small blue circles, of all line intersections (the lines are automatically hidden from the display for an ease of view, as shown in Fig. 4 (c)). Each intersection is assigned an unique numerical ID. Some, but not all, of these intersections correspond to physical wall corners.

The next step is for the user to (1) select which ones of the intersections do correspond to wall corners, (2) select whether two nearby corners are joined by a wall, and (3) associate corners and walls to individual spaces (rooms or open areas such as corridors or halls). Note that complex spaces can be conveniently subdivided into smaller spaces as shown in Fig. 5 (a). An example with a diagonal wall is shown in Fig. 5 (b). Note that corners #32 and #53 are generated as the intersections of proper wall lines (i.e., lines containing actual wall segments) with "ghost wall" lines, as explained above. For example, a horizontal ghost wall line (created by a Shift-click on the map) intersects the vertical wall line at #32.

The corner selection and wall association step is accomplished as follows. Each space is visited in turn. At each space, the user clicks on the line intersections (the small blue circles) that correspond to physical wall corners within that space. The color of the selected wall corners turns red, and their associated IDs are displayed on the map (Fig. 4 (d)). These wall corners are sorted in the clockwise order, and listed in the control panel of the interface. In addition, all possible walls joining adjacent corners are also listed in the same panel. For example, in Fig. 4 (e), after the user selects corners #18, #22, #42, #46, they are ordered as (#18, #42, #46, #22), and all possible walls connecting adjacent corner pairs are displayed. These are: (#18, #42), (#42, #46), (#46, #22), and (#22, #18) (not shown in the figure due to space limitation). The user then simply clicks on the corner pairs that correspond to actual walls, which are then displayed as red segments. In this case, the room has a closed contour (except for door openings), hence all corner pairs are selected.

As another example, consider the open space ('LOBBY') shown in Fig. 5 (c). Its fairly complex layout is divided into a number of smaller spaces, one of which is defined by the wall corners (#33, #47, #46, #53, #56, #35). Only the following corner pairs are joined by a wall: (#33, #47), (#47,#46), (#53,#56). Note that the remaining corner pairs ( (#46, #53), (#56, #35), (#35, #33)) are not selected, signifying that the spaces between them are open.

In order to trace an entrance door of a space, the user first defines the whole wall containing the door as described above (as opposed, for example, to defining two wall segments at either side of the door). Once the wall segment has been determined, one can define the endpoints of the door segment, by Shift-clicking on the appropriate locations on the wall segment. For example, in Fig. 4 (f), the user specifies two entrances along the wall connecting the corner-pair (#18, #42), as well as two entrances along the wall of (#46, #22). Two endpoints are automatically stored in a list (separate from the wall corners list), and a new entity ("entrance") is defined, joining the two endpoints. The user concludes the task of tracing a space by providing its name (e.g., a room number), and by selecting its space type from a pull-down menu. The currently supported types include: room (default), corridor, restroom, staircase, elevator.

*3.1.2 Spatial Features Representation.* The floor plan tracing process described above produces a spatial information hierarchically organized in terms of *spaces*. Each space is characterized by a set of *wall corners* and possibly *entrance corners*, where pairs of adjacent wall corners may or may not be joined by a wall. We store this information in a *sim* file. Our sim format is inspired by the Polygon File Format (PLY), which is used to represent 3D objects as lists of flat polygons. A PLY file contains a list of vertices and a list of polygons, where each polygon is defined as an ordered list of vertex IDs. A sim file contains a list of wall corners and a list of entrance corners. Each space is assigned a list of wall corner IDs and a (possibly empty) list of entrance corner IDs. Additionally, sim allows one to specify whether two wall corners in the list should be connected by a wall, or not (implying an empty space between these corners).

A space is represented in the following format: `{id type name num_corner wall_corner_indices walls entrances}`. For example, `{s2 0 217 5 1 3 27 19 12 1 1 1 1 0 e2 3 1 2}` means that the space's ID is s2, its space type is 0 (meaning a room by default), its name is 217, and it has 5 corners whose indices, sorted in clockwise order, are (#1, #3, #27, #19, #12). The next sequence of binary values (`1 1 1 1 0`) indicates that there are walls connecting the corner-pairs (#1, #3), (#3, #27), (#27, #19), and (#19, #12); but there is no connection for (#12, #1). The last sequence with four entries (`e2 3 1 2`) denotes that there is an entrance with an identifier of e2 along the wall with index 3 (i.e., the third wall in the list: (#27, #19)). The endpoints of this entrance are (#1, #2), where these IDs refer to the list of entrance corners. Additional entrances to the same space can be listed as additional quadruplets of entries at the end of the list. Note

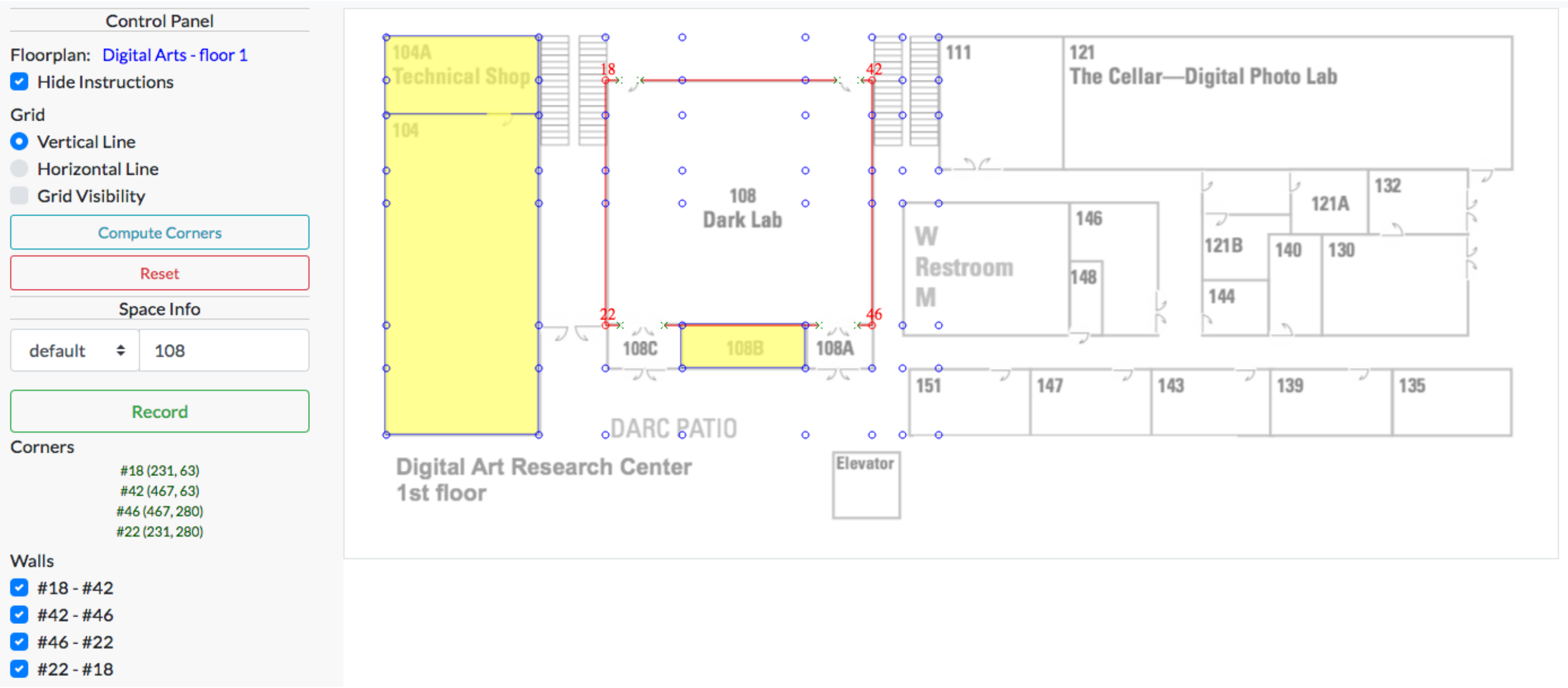

**Figure 3: The toolkit GUI includes a control panel and a canvas. Possible wall corners (line intersections) are rendered as blue circles. Actual wall corners (line intersections) are rendered as red circles. Red segments represent wall segments. Gaps with endpoints marked by a green x denote entrances within wall segments. Yellow polygons with blue border show spaces for which a tracing was completed. Gaps in a blue border represent entrances to a space.**

that wall corners and wall segments can be re-used for different adjacent spaces.

*3.1.3 Conversion to GeoJSON.* A sim file can be easily converted into other formats. The Map Conversion toolkit contains a converter into GeoJSON, a popular format for representing spatial information [2]. The GeoJSON *features*, representing segmented spaces or annotated objects, consist of a set of *properties* represented as a (key, value) mapping and a geographical *geometry* represented as a polygon (one of the geometric primitives in the GeoJSON format). A feature's properties include name, encoded color, height, and distance from the ground level. The geometry information contains the coordinates (lat, long) of the polygon's vertices. Multiple features are hierarchically grouped into a *Feature Collection* object. The GeoJSON generator in the Map Conversion toolkit generates the *Feature Collection* object automatically from the *sim* structure. Similarly, objects of interest (Sec. 3.2) are annotated and then represented as cuboids. Fig. 6 shows an example of conversion from sim spaces to GeoJSON Feature Collection. The Mapbox GL JS engine renders a 3D map view by extruding each feature in such collection.

*3.1.4 Geo-registration.* Corners in sim are defined in terms of (x, y) screen coordinates. The conversion to (lat, long) geodetic coordinates for GeoJSON representation is performed as follows. We first determine the geodetic coordinates of at least four corners, chosen from the building's external walls. This is easy to do if, for example, the contour of the building under consideration is visible in a web application such as Google Map or Apple Map, and the location of the selected corners can be identified in this contour (note that these applications return the WGS84 geodetic coordinates of selected locations). The geodetic coordinates of these points are then converted to Universal Transverse Mercator (UTM) coordinates using standard formulas. The UTM system is based on a conformal projection, and thus produces little distortion for small areas. Next, we determine a collineation (homographic) transformation between the (x, y) screen coordinates of the wall corners and the UTM coordinates of the same points. The collineation matrix can be found using Direct Linear Transformation [21]. The same collineation is then used to transform the (x, y) coordinates of all remaining corners into UTM coordinates, which are then converted into geodetic coordinates.

Some examples of an end-to-end conversion from floor plan images to GeoJSON files, shown as pop-ups over OpenStreetMap using the MapboxGL JS engine, are presented in Fig. 7. Note that in the conversion from sim to GeoJSON, staircases have been represented as three adjacent rectangles of different heights, colored in green, while elevators are shown colored in blue. Some advantages of using our Map Conversion toolkit as compared to other web-based drawing interfaces are highlighted in Fig. 8, which compares the results using MapBox Studio (left) and our Map Conversion toolkit (right). Mapbox studio, like other drawing interfaces such as Google Map, doesn't allow one to trace a floor plan image. Individual walls need to be copied by hand, often resulting in geometric errors such as incorrect spacing or orientation. In addition, when shapes are drawn manually by hand, connectivity errors may occur (see an inset of Fig. 8). Our strategy of first defining a line grid, and then selecting corners from the line intersections, ensures that co-planar walls are represented by collinear segments, and that connected wall corners remain connected.

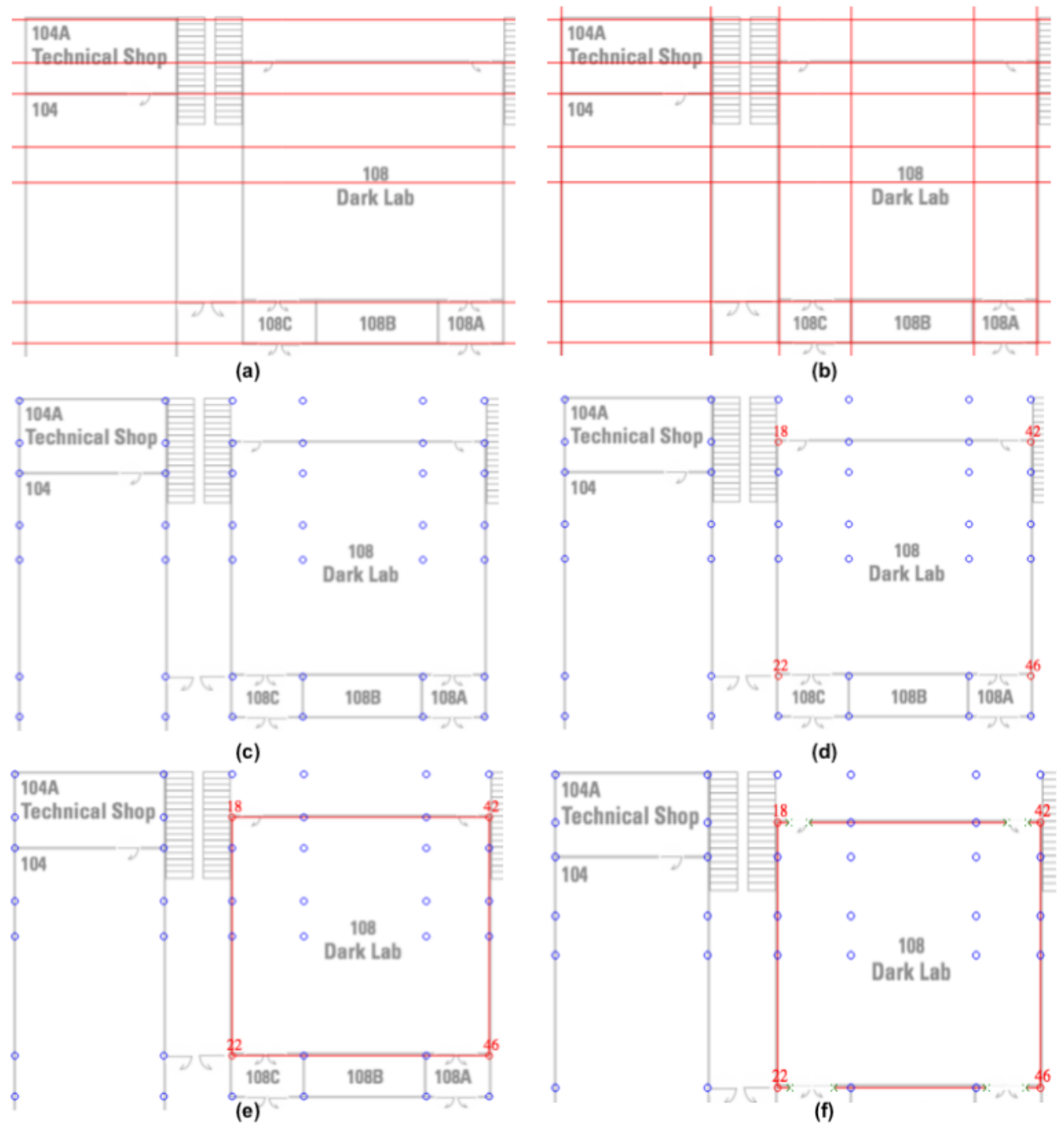

**Figure 4: A typical Map Conversion workflow. (a) First, horizontal grid lines are generated via Shift-clicks. (b) Vertical grid lines are generated next. (c) All line intersections (possible wall corners) are automatically computed and displayed. (d) The user selects corners #18, #22, #42, #46 for the boundary of a space (Room 108). (e) The user then selects the walls connecting the corner pairs (#18, #42), (#42, #46), (#46, #22), and (#22, #18). (f) Finally, the user defines two entrances along the wall of (#18, #42) and two entrances along the wall of (#46, #22).**

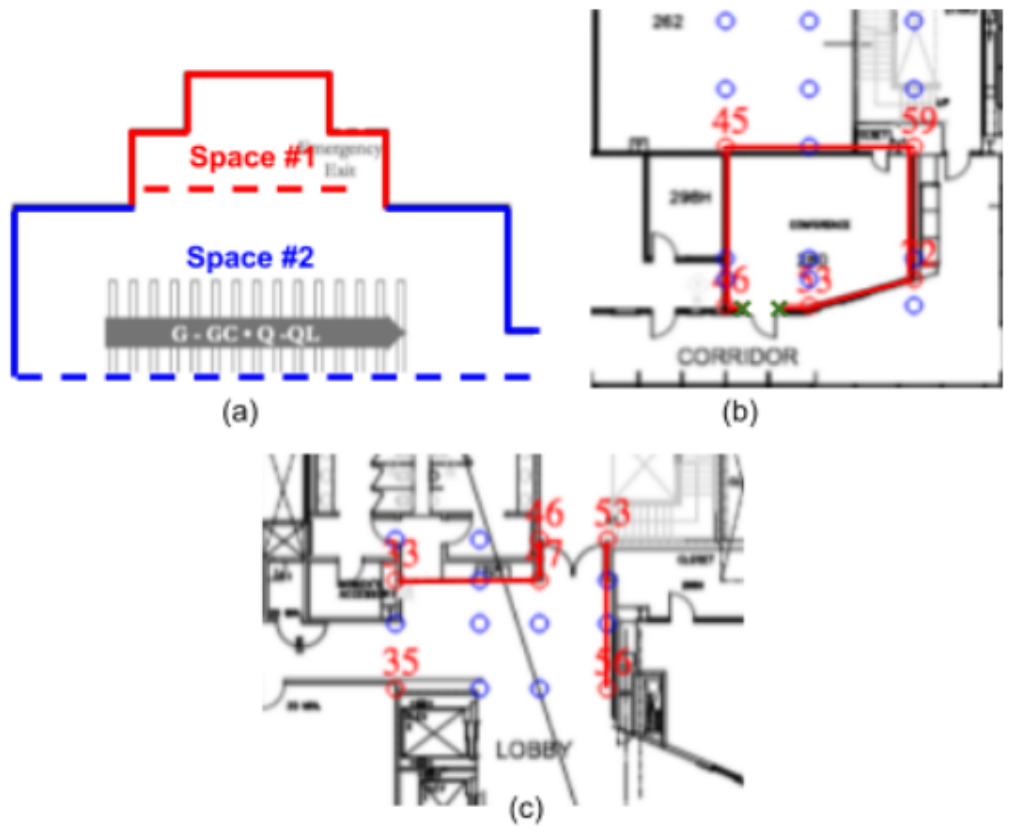

**Figure 5: (a) A complex space can be divided into multiple spaces to facilitate tracing. (b) An example of a traced room with a diagonal wall. (c) Tracing a "segment" of an open space. Note that several wall corners are not linked by walls.**

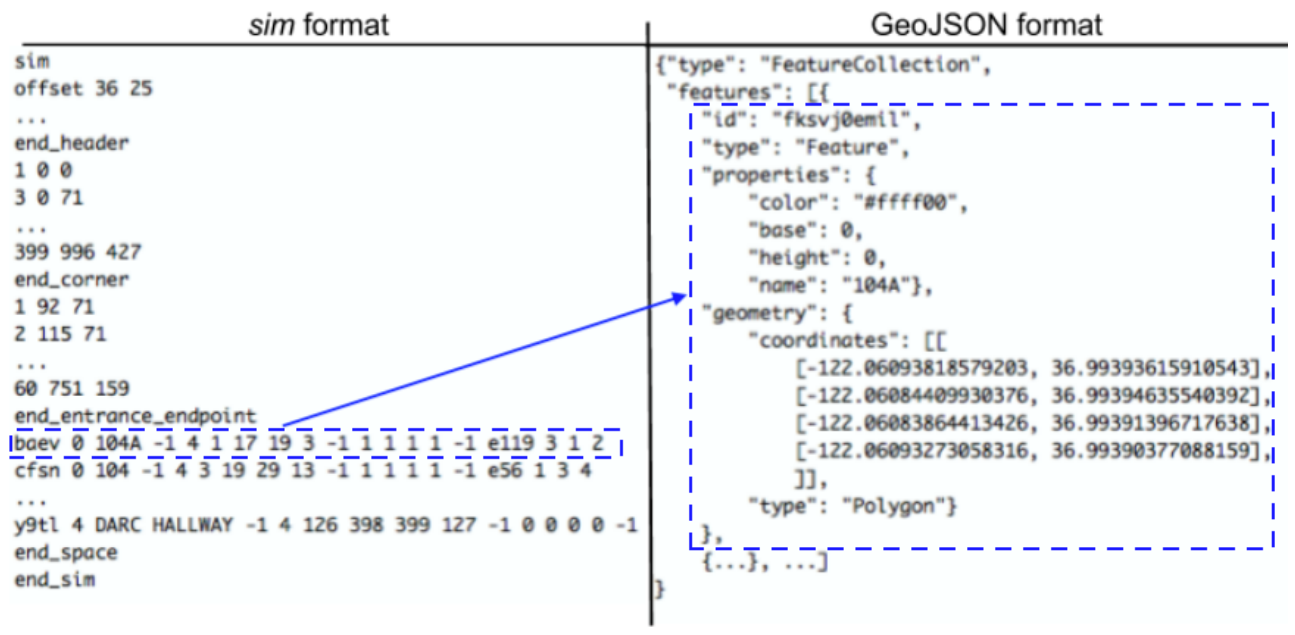

**Figure 6: An example of conversion from sim to GeoJSON. A room, described by one row in the sim file, is represented as a feature in GeoJSON.**

## 3.2 Map Population Toolkit

The Map Population toolkit allows one to insert 3D objects, such as furnitures, into a GeoJSON representation of a floor plan. Objects are extracted from a 3D scan of the environment, and represented as boxes. We use Occipital's Structure sensor, which has a RGB-D camera and software for registration and stitching of multiple 3D point clouds into one mesh, stored in a PLY format. The workflow is organized in a sequence of stages (orientation, rectification/rescaling, segmentation), as described next. We will assume that a room with four walls has been scanned in its entirety; the same mechanism can be extended to the case of partial scans, or for different types of spaces (e.g., corridors).

*3.2.1 Mesh Orientation.* Our first step is to orient the mesh acquired by the 3D scanner with the floor plan. The Structure sensor produces a mesh with its Y-axis vertical (as measured by the sensor's accelerometer), but with an arbitrary orientation of the X-Z plane. We would like to re-orient the mesh (rotate it around the Y-axis) such that the walls of the room are aligned with the X and Z axes. (These axes will then be mapped to the axes of the 2D floor plan). We first select all vertices in the mesh whose normal vector is approximately orthogonal to the Y-axis (i.e., corresponding to vertical surface elements). For each such vertex, we compute the angle formed by its normal and the Z-axis. Peaks in the histogram of these angles reveal the orientation of the main walls. For example, in the case shown in Fig. 9 (a), a peak is found at 127°, corresponding to the orientation of the longer walls. The whole mesh is then re-oriented by rotation around the Y-axis by the opposite angle (Fig. 9 (b)).

*3.2.2 Mesh Rectification/Rescaling.* Due to errors in data acquisition, registration or stitching, the geometry of 3D scans of environments is often inaccurate. In particular, wall scans are sometimes not planar, or walls appear not to intersect at 90°. This may affect the registration of the environment with the floor plan. We correct for global errors using the following simple procedure. We first identify the four walls in the acquired mesh. To do this, we project the vertices of the re-oriented mesh onto the X-Z plane (Fig. 10 (e)). We then select the vertices with Z-coordinate in the top quartile, and run the RANSAC algorithm [8] to find a robust line fitting (Fig. 10 (a)). This line represents the top wall. We repeat the same procedure for all sides (Fig. 10 (b-d)), resulting in four lines in the X-Z plane. An example of the result is shown in Fig. 10 (e). From this figure, it is clear that, due to artifacts of scanning, the walls do not appear to intersect at 90°. We then find the axis-parallel rectangular box that best approximates the quadrilateral formed by the line intersections (Fig. 10 (f)). The collineation (homography) that brings this quadrilateral's vertices into the corners of this axis-parallel rectangle is computed. The mesh can then be rectified by applying the same collineation to the (X, Z) coordinates of all vertices in the mesh.

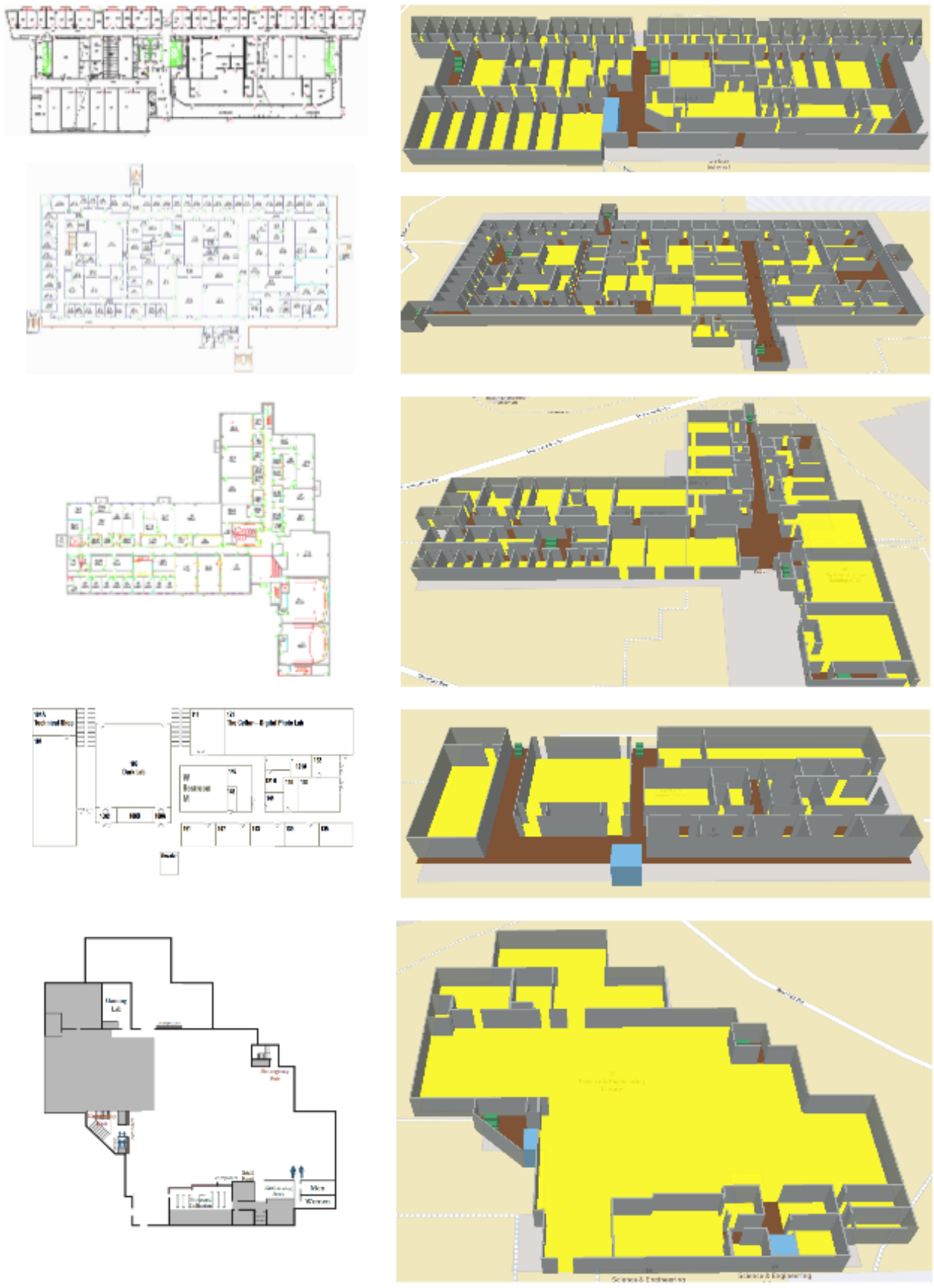

**Figure 7: Examples of application of our Map Conversion toolkit.**

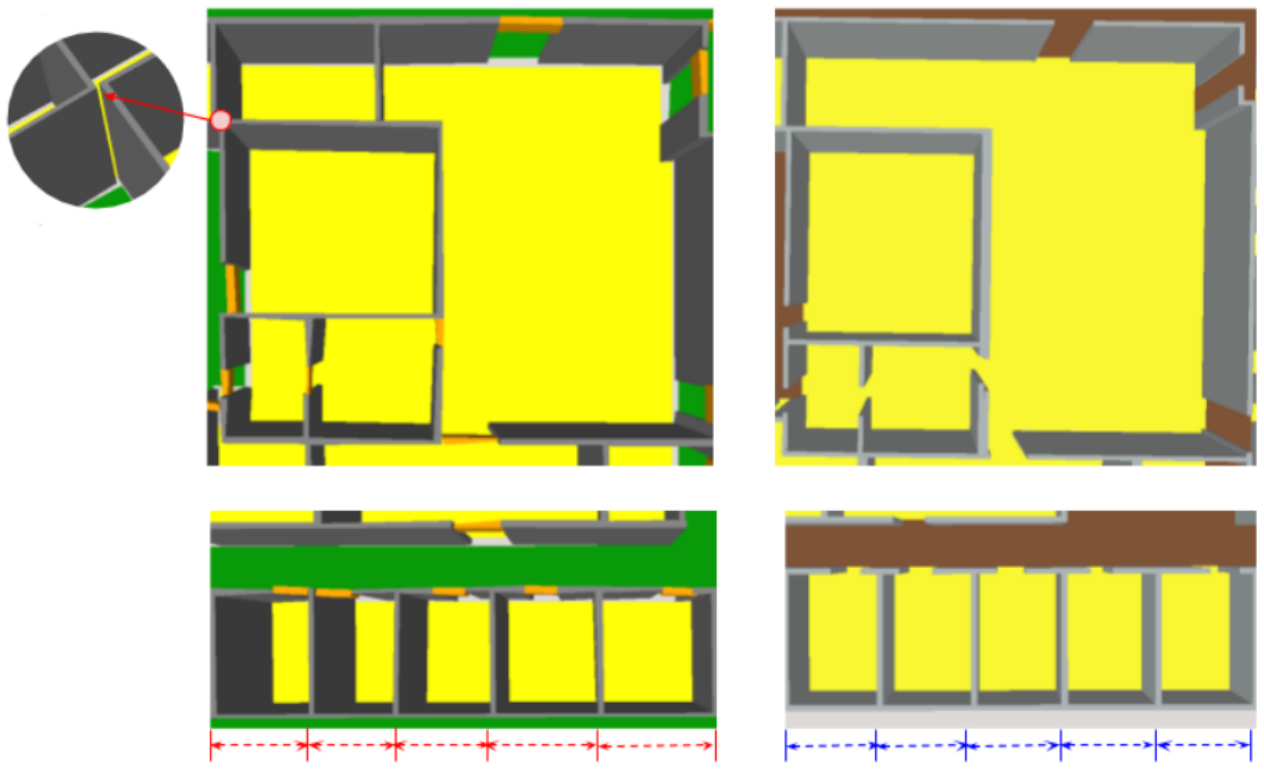

**Figure 8: The same floor plan extracted and saved as a GeoJSON file using Mapbox Studio (left) and our Map Conversion toolkit (right).**

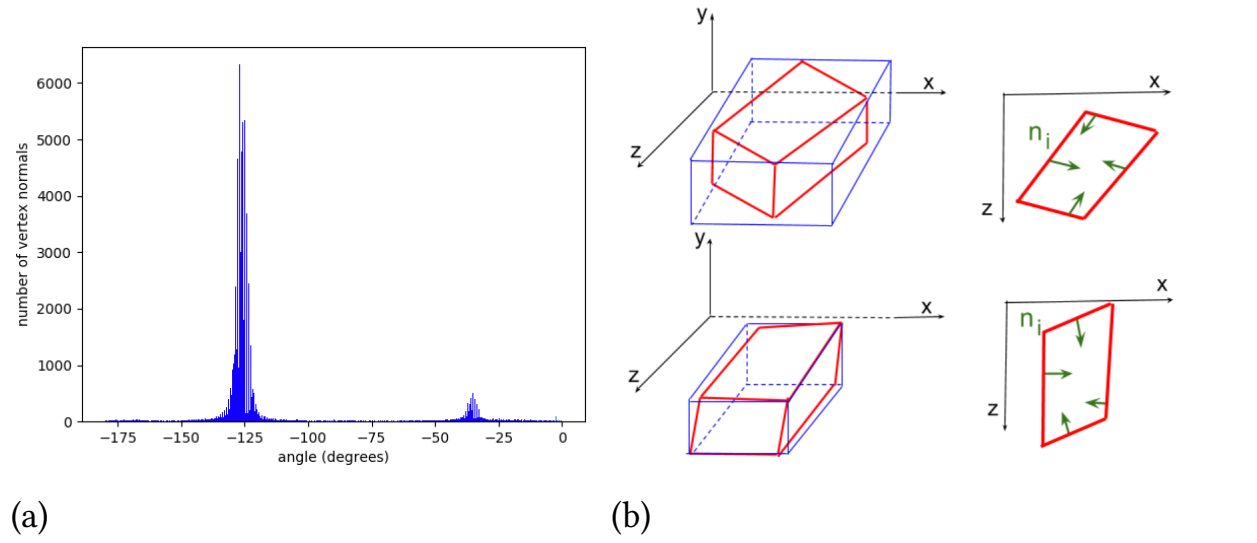

**Figure 9: A simple example of re-orientation. The histogram of the angles between surface normals and Z-axis is shown in (a). The top row of (b) has the original shape, with a representation of the faces' normals. After re-orientation, the longest segment is aligned with the Z-axis.**

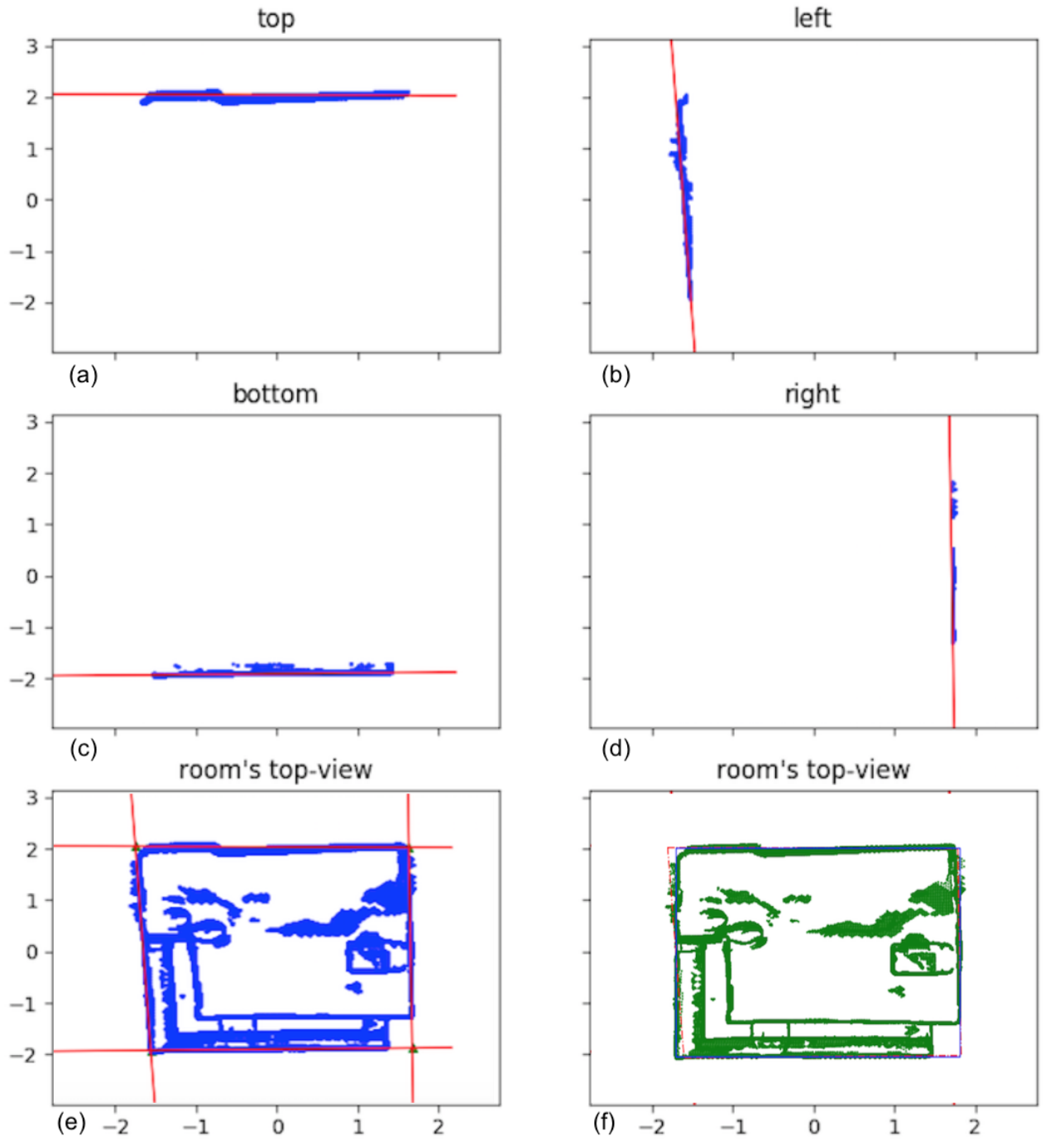

**Figure 10: Individual walls are identified by linear fitting of the vertices of the re-oriented mesh, projected onto the X-Z plane (a - d). The resulting quadrilateral (e) is transformed into the best-fitting axis-parallel rectangle (f), and the same transformation is applied to the (X, Z) coordinates of all vertices in the mesh.**

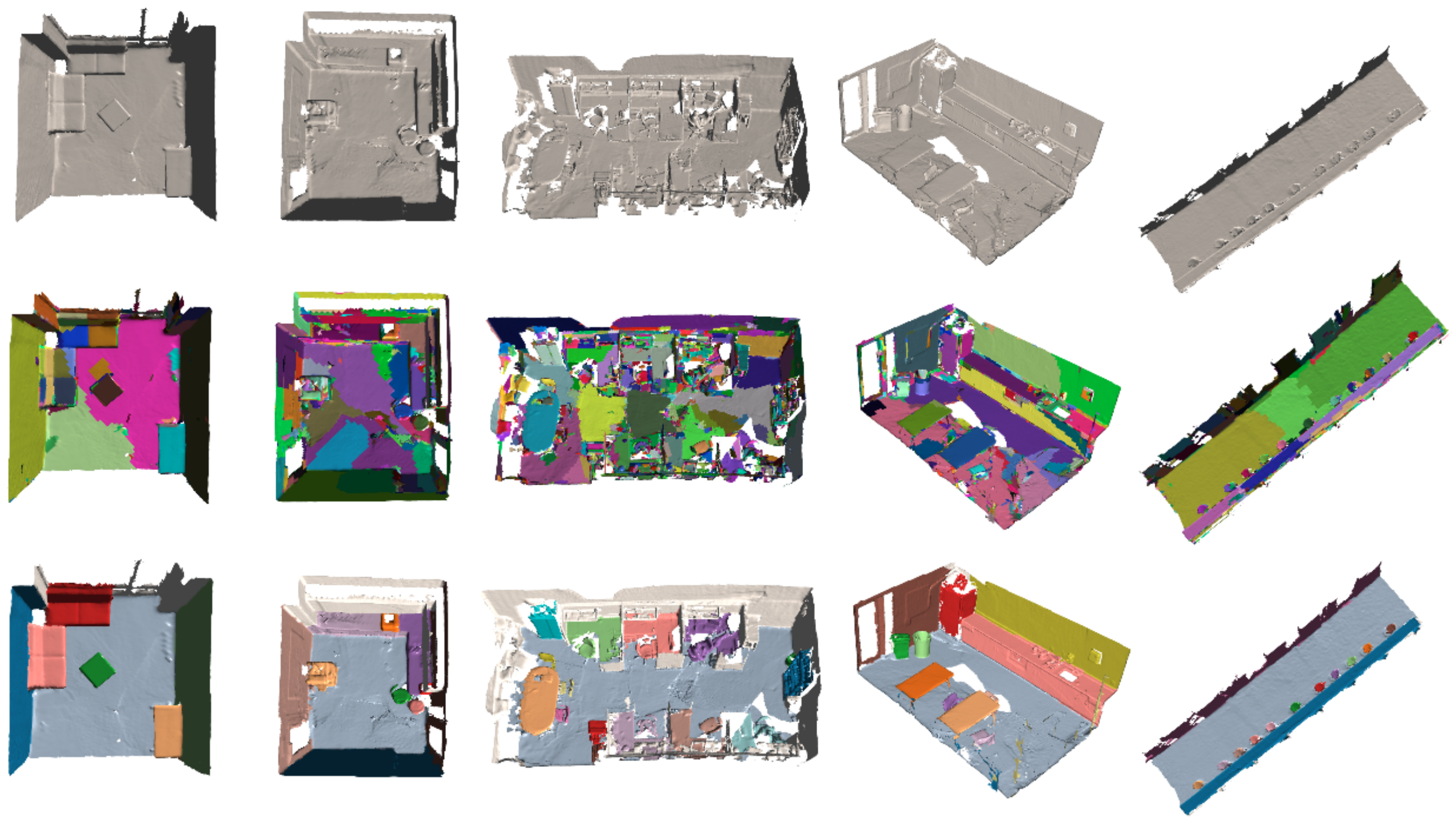

**Figure 11: Top row: 3D scans of indoor environments using Occipital's Structure sensor. Center row: automatic super-pixelation of the meshes using the algorithm of [7]. Bottom row: segmentation of individual objects using the web toolkit described in [3].**

In order to register the resulting mesh with the floor plan, we first visually determine the correct orientation of the mesh. From the Sec. 3.2.1, our re-orientation procedure aligns the longer walls with the Z-axis. However, this may not be the actual orientation of the space in the floor plan, and an additional rotation by $\pm 90^\circ$ or $180^\circ$ may be required. Finally, we find the offset between the (x, y) coordinates of one corner of the room in the floor plan, and the (X, Z) coordinates of the corresponding corner of the rectangular bounding box, as well as the two scale factors that ensure that the mesh correctly fits the room in the floor plan.

*3.2.3 Object Segmentation.* The final goal of the Map Population toolkit is to extract objects of interest from the 3D scan, correctly dimensioned and registered with the floor plan, and insert them into the GeoJSON feature collection, in the form of cuboids placed on the ground. We employ the following semi-automatic procedure for the task of segmenting objects from the re-oriented and rectified mesh. First, we use the algorithm described in [7] to generate "super-pixels", which in this case correspond to a connected sets of mesh facets with a similar orientation. A super-pixel thus represents an approximately planar surface patch. We then use the web-based toolkit described in [3] to manually select all super-pixels corresponding to each identified object. Examples of automatic super-pixelation and manual segmentation (assignment of super-pixels to individual objects) are shown in Fig. 11. Finally, for each annotated object, we compute the bounding box of the associated sub-mesh, and assign it a name and a color for display. Since the mesh was already correctly registered with the floor plan, adding the spatial description of this bounding box to the same GeoJSON file is trivial. Some examples of map population with individual objects are shown in Fig. 12.

## 4 CONCLUSIONS

We have presented SIM, a toolbox for the generation of vectorized representation of floor plans and of room content. The Map Conversion toolkit allows one to quickly and accurately trace the layout of a floor plan, and generate a geo-registered GeoJSON file that can be visualized interactively. The Map Population toolkit allows one to populate spaces with objects such as furnitures. Rather than manually measuring the size and location of these items, the user can simply take a 3D scan with a RGB-D camera, and easily segment out individual objects using the toolkit. The 3D mesh is rectified and registered with the GeoJSON representation of a floor plan so that the objects are automatically placed in their correct location on the map. We are planning to make the SIM toolbox available to the public as a web application in the near future. We believe that SIM represents an innovative and useful tool for the production of vectorized maps, and that its simplicity of use may appeal to both practitioners and casual users.

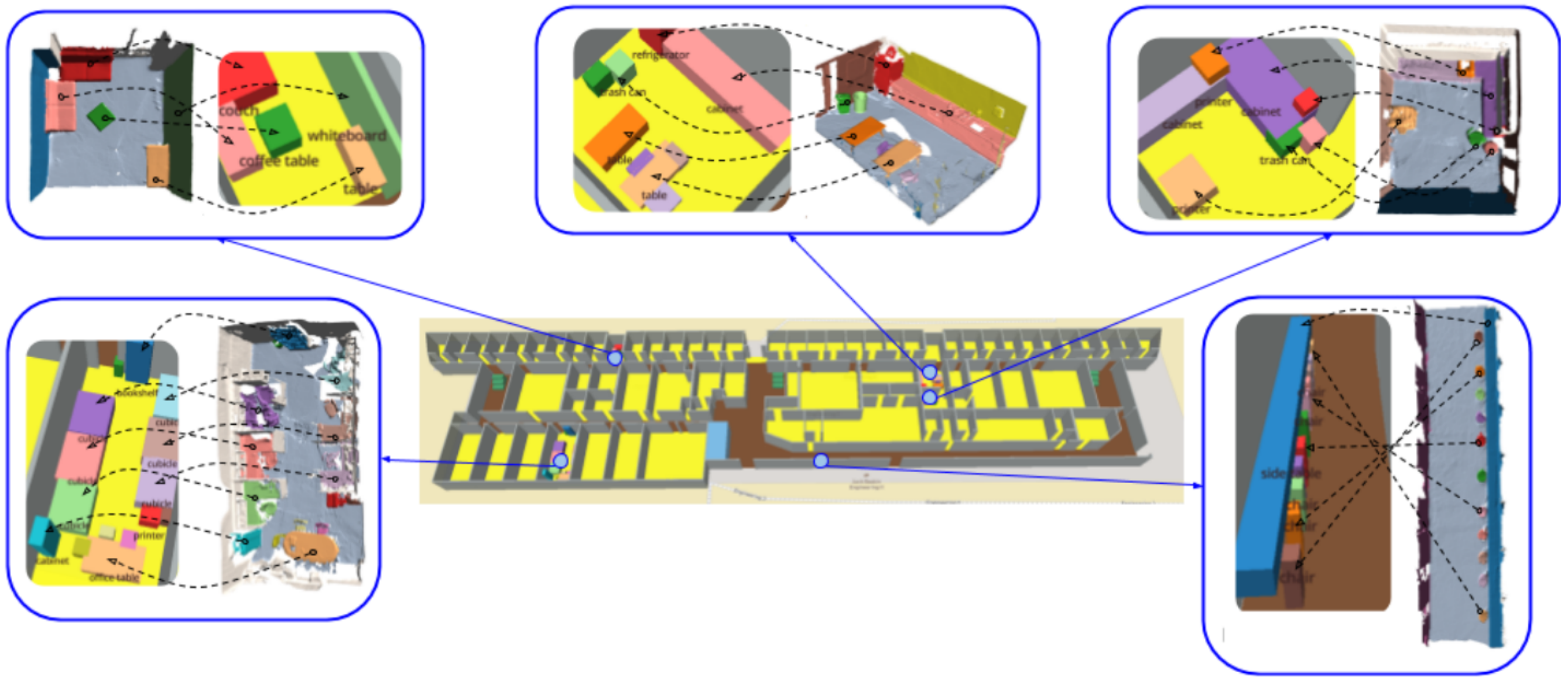

**Figure 12: The bounding boxes of individually segmented objects are placed in the GeoJSON file containing the building's floor plan. The entire building with its objects of interest is displayed over OpenStreetMap.**